\documentclass[12pt]{article}

\usepackage{a4wide}
\usepackage{graphics}
\usepackage{amsfonts}
\usepackage{amstext}
\usepackage{amsmath}
\usepackage{cite}
\newcommand{\Section}[1]{\setcounter{equation}{0}\section{#1}}
\def\Tr{\text{Tr}}
\def\ln{\text{ln}}

\def\0{\hbox to 12pt{\hss$0$\hss}}
\def\+{\hbox to 12pt{\hss$+$\hss}}
\def\-{\hbox to 12pt{\hss$-$\hss}}
\def\E#1{\langle#1\rangle}

\begin{document}       
\title{
Spatial Particle Condensation for an Exclusion Process on a Ring
}

\author{
\vspace{5mm}
N. Rajewsky,{\footnote {\tt e-mail: nr@eds3.rockefeller.edu}} \,
T. Sasamoto\,{\footnote {\tt e-mail: sasamoto@monet.phys.s.u-tokyo.ac.jp
 (Corresponding author.) }}\,\,
and E. R. Speer\,{\footnote {\tt e-mail: speer@math.rutgers.edu}}\\
{\it $^{*}$Center for Studies in Physics and Biology,}\\ {\it The Rockefeller University,
1230 York Avenue,}\\
\vspace{5mm}
{\it New York, NY 10021, USA.}\\
{\it $^{\dag}$Department of Physics, Graduate School of Science,}\\
{\it University of Tokyo,}\\
\vspace{5mm}
{\it Hongo 7-3-1, Bunkyo-ku, Tokyo 113-0033, Japan.}\\
{\it $^{*\,\ddag}$Department of Mathematics,}\\
{\it Rutgers University,}\\
{\it New Brunswick, New Jersey 08903, USA.}}

\date{} 

\maketitle
\begin{abstract}
We study the stationary state of a simple exclusion process 
on a ring which was recently introduced by Arndt {\it et al}
\cite{AHR98-3,AHR98-4}. This model
exhibits spatial condensation of particles. It has
been argued \cite{AHR98-3,AHR98-4} that the model has 
a phase transition from a ``mixed phase'' to a
``disordered phase''. However, in this paper exact
calculations are presented which, we believe, show that 
in the framework of a grand canonical
ensemble there is no such phase transition. An analysis of 
the fluctuations in the particle density
strongly suggests that the same result also holds
for the canonical ensemble.

\end{abstract}

\Section{Introduction}
\label{intro}
One-dimensional driven diffusive systems have 
attracted the interest of researchers in a wide variety of 
fields, including mathematics, physics, chemistry and biology
\cite{L,L2,Sp,SZ}. One of the most basic of these models is the asymmetric
simple exclusion process (ASEP), which already exhibits a rich
behavior. Within physics, for example, the ASEP
has been used to study boundary induced phase transition
\cite{Krug}, spontaneous symmetry breaking 
\cite{EFGM}, phase separation \cite{EKKM} and shocks 
\cite{DJLS,DLS}.
As one of the few models
in non-equilibrium physics which is analytically accessible, it has 
served as a laboratory to test basic concepts
of statistical mechanics for far from equilibrium systems. 
Moreover, its simplicity has led to applications in such diverse fields 
as the kinetics of biopolymerization \cite{biopoly}, vehicular
traffic \cite{procs97,ERS} and, most recently, sequence alignment
problems in biology \cite{bundschuh}.

In this paper, we study a related exclusion process which
 exhibits a spatial condensation of particles.
This model was introduced by Arndt, Heinzel and Rittenberg 
 \cite{AHR98-3,AHR98-4} and will be called AHR~model below. 

In the AHR~model two types of particles, called {\it positive} 
and {\it negative}, occupy the sites of 
a periodic one-dimensional lattice of length $L$.
The particles  are subject to hard-core exclusion, so
that there are three possible states at each site:
empty, occupied by a positive particle, or
occupied by a negative particle. 
Positive particles hop at rate $\alpha$ 
into empty spaces (holes) to their immediate right, and 
negative particles at the same rate into empty spaces 
to their left.  Adjacent positive and negative particles also 
exchange positions with asymmetric rates $p_R$ and $p_L$. 
Specifically, the rules are
\begin{equation}
\begin{matrix}
   & \+\0\quad\rightarrow\quad\0\+  &\text{with rate} & \alpha, \\
   & \0\-\quad\rightarrow\quad\-\0  &\text{with rate} & \alpha, \\
   & \+\-\quad\rightarrow\quad\-\+  &\text{with rate} & p_R, \\
   & \-\+\quad\rightarrow\quad\+\-  &\text{with rate} & p_L; 
\end{matrix}
\end{equation}
we assume that  $p_R,\alpha > 0$ and $p_L\geq 0$. 
A key parameter in the model is ratio of $p_L$ to $p_R$, which 
we denote by $q$:  $q=p_L/p_R$. Our notation here is {\it not} that 
of \cite{AHR98-3,AHR98-4}, where $q$ denotes the inverse of this ratio; we
adopt this change to agree with the standard notation in the 
theory of basic hypergeometric functions 
(see e.g. \cite{GR}).  
 Finally, we normalize our time scale by setting $p_R=1$, so that the rates
for exchange of positive and negative particles are $1$ and $q$. 
Note that the numbers $N_\pm$ of positive and negative particles are
constants of the dynamics; we will write $\rho_\pm=N_\pm/L$ for the particle
densities. 

We remark that
when $p_L=0$ and $p_R=\alpha=1$ the model corresponds
to a two-species ASEP (with ``first'' and ``second class'' particles) 
which has been solved in
\cite{DJLS}. The system with no holes ($N_++N_-=L$)
corresponds to the partially asymmetric exclusion
process on a ring, for which  the stationary 
measure is trivial (all configurations are equally likely).

The stationary state of the AHR~model displays a rich dynamical
behavior, which was extensively investigated in \cite{AHR98-3,AHR98-4}.
Two  methods were used:
Monte Carlo simulations were carried out in the {\it canonical ensemble}
(CE), in
which the numbers $N_\pm$ of particles are fixed, and analytic
calculations (still in finite systems) using a quadratic algebra
were carried out in a {\it grand canonical
ensemble} (GCE), in which the densities are controlled 
by fugacities $\xi_\pm$ and can fluctuate. 
These results were then extrapolated to infinite volume.
Most of the work was done in the case of equal densities 
($N_+=N_-$ or $\zeta_+=\zeta_-$), which is the only case we 
will consider here.

These investigations strongly suggest the existence of
three different phases for the model, 
differing in the nature of typical spatial 
particle configurations (see Figure~1 for space time representations).
We summarize the discussion of \cite{AHR98-3,AHR98-4} and the nomenclature
adopted there.
 For $q>1$ the system is in the ``pure
phase'': typical configurations contain three blocks, one each of holes,
positive particles, and negative particles.  
In the thermodynamic limit ($L \rightarrow \infty$) 
the current vanishes.  The system in this phase is closely related to
a class of models studied in \cite{EKKM}.
 For $0\leq q<1$ the phase depends on the density $\rho=\rho_+=\rho_-$. 
 For relatively small densities, density profiles of both species 
are uniform and there is no spatial condensation (the ``disordered
phase''). For larger densities one observes the
formation of a condensate 
made of $+$ {\it and} $-$ particles (with
some holes at the boundaries); a block of holes,
occasionally traversed by $+$ or
$-$ particles, occupies the
rest of the system (the ``mixed phase''). 
At the transition between the mixed and disordered phases
the Monte Carlo simulations suggest 
that the infinite volume current $J$ of $+$ particles
increases monotonically from zero to a value $(1-q)/4$
as the the density $\rho$ increases from zero to a 
certain value $\rho_c$, and then takes the constant value
$(1-q)/4$ for $\rho_c<\rho<1/2$
(see Figure 8 of \cite{AHR98-4}). 
This indicates that the function $J(\rho)$ is 
non-analytic---in fact, the simulations suggest it is not 
differentiable---at a certain
value of density $\rho_c$. 
The same apparent non-analyticity 
is seen in plots, derived from calculations in the GCE,
of $J$ as a function of $q$ at fixed $\rho$ 
(Figures~1 and 2 of \cite{AHR98-4}).
This would correspond
to a ``phase transition'' in the language of equilibrium
physics.  

In fact, however, we believe that our analysis shows that 
these numerical results are misleading. 
Our main result is an explicit exact formula
for the current $J$ in the infinite system in the framework
of a GCE. This computation was possible
because the AHR model is closely related to the 
one-dimensional partially asymmetric simple exclusion 
processes, for which recently analytic exact 
expressions for various expectation values have been computed 
\cite{PASEP,PASEP2,BECE}.
It turns out that if $q<1$ then the current $J(\rho)$ 
is analytic everywhere, although
the derivative of $J(\rho)$ changes very rapidly as $\rho$ passes 
$\rho_c$ when $q$ is close to one. 
This strongly suggests that in the GCE, 
there is no phase transition.  
For this reason, we will here speak of the 
{\it mixed regime} and the 
{\it disordered regime}, and refer to the union of these regimes 
as the {\it mixed/disordered phase}.

Now, is this also true for the CE?
Based on the observation that the density fluctuations 
in the GCE go to zero in the thermodynamic limit,
we believe that the answer to this question is yes,
although the argument is not rigorous (See section~5 for a more 
detailed discussion.)

The existence of a phase transition between the pure and 
the mixed/disordered
phase seems uncontroversial \cite{AHR98-3,AHR98-4} and will not be discussed in
the following.  Therefore, $q$ is normally chosen to satisfy 
$0\leq q< 1$ from now on, although we will occasionally 
consider $q=1$ as a special limiting case. 

This paper is organized as follows. In section~2, we present the
basic structure of the exact solution and define the GCE and CE.
The current and density in the GCE are computed in section~3. 
An analysis of the model when  $q$ is close to one
is carried out in section~4. In section~5 we  discuss
to what extent  the results for the GCE are valid
for the CE. The final section contains
our conclusions.

\Section{Exact solution of the stationary state}

The stationary state 
of the AHR~model can be constructed by applying
the so-called matrix product ansatz (MPA) \cite{HN,DEHP}; this is the method
used for the grand canonical calculations in 
\cite{AHR98-3,AHR98-4}. Since its original
application to the ASEP in
\cite{DEHP}, various exact results for exclusion
processes have been achieved via the MPA; see for example
\cite{PASEP} and references therein. 
Since the method is now well known, we will
skip a detailed presentation of the formalism and
even omit the proof that the particular version presented here
correctly determines the stationary state.

The MPA expresses the probability of any 
configuration in the stationary state of the AHR~model, up to an overall
normalization,  as a trace of a
product of certain matrices.  
Thus, for example, the probability $P$ of the 
configuration $+-00+$ can be written in the form 
$P(+-00+)= Z^{-1}\Tr(DEAAD)$; here $Z$ is an ensemble-dependent 
normalization constant which will be 
discussed shortly.
 The matrices $D$, $E$ and $A$ are
assigned to the local states $+$, $-$ and $0$, respectively, and 
are required to satisfy the
following algebraic conditions, which guarantee stationarity 
of the resulting state:
\begin{gather}
   DE - q ED = \zeta (D+E) , \notag\\
  \alpha DA       = \zeta A ,     \notag\\
  \alpha AE       = \zeta A .  
  \label{alg}
\end{gather}
Here the number $\zeta$ is arbitrary; a natural choice, which we
adopt from now on, is 
 \begin{equation}
    \zeta = 1-q. 
   \label{zeta}
 \end{equation}
The conditions (\ref{alg}) 
are closely related to those for the matrices of the MPA 
for the partially asymmetric simple exclusion processes  
\cite{PASEP,PASEP2,BECE}.

It is convenient to introduce the new parameter
\begin{equation}
a
=
-1+\frac{1-q}{\alpha}.
\end{equation}
In terms of $a$, explicit representations for
the matrices $D$, $E$ and $A$ which satisfy (\ref{alg}) are
given by
\begin{align}
D
&=
\begin{bmatrix}
1+a     & \sqrt{c_1} & 0          & 0          & \cdots  \\
0       & 1+aq       & \sqrt{c_2} & 0          &         \\
0       & 0          & 1+aq^2     & \sqrt{c_3} &         \\
\vdots  &            &            & \ddots     & \ddots  \\
\end{bmatrix},
\notag\\
E
&=
D^T,
\notag\\
\noalign{\noindent and}
A&=|0\rangle \langle 0|,
\label{de-aw}
\end{align}
where the superscript $T$ indicates the transpose,
\begin{equation}
\label{cn-aw}
c_n
=
(1-q^n)(1-a^2q^{n-1}),
\end{equation}
$\langle 0|=[1\ 0\ 0\ \ldots\ ]$, and  $|0\rangle=\langle0|^T$. 

Recall that we take $0\leq q <1$, and certainly $a>-1$.  Thus if $a<1$ then
$c_n$ is positive for any $n ~(\geq 1)$, 
and the roots in (\ref{de-aw}) are real. If $a\ge1$ then
the roots may be imaginary; moreover, $c_n$ can vanish for special values
of $a$ and $n$, leading to finite dimensional representations. The case
$a\ge1$ requires special treatment in other ways, which we will mention
briefly in section~3.  In the future, unless special mention is made, we
assume that $-1<a<1$; this is the interesting case since 
the transition between the mixed and disordered regime occurs in this
region (see Table~1 of \cite{AHR98-4}; our results below show that this
transition in fact occurs for $a<0$).

Since the matrices
in (\ref{de-aw}) are infinite dimensional, it is not {\it a priori}
clear that the trace of an arbitrary product of these matrices
exists. However, $A$ is 
a projector of rank one, so that
the trace of any product of these matrices will
exist as long as there is at least one $A$ matrix in the product or,
equivalently, as long as there is at least one hole in the
corresponding configuration. We will consider here only such well defined 
traces since, as
mentioned in the introduction, when there are no holes the model has
a trivial steady state.

For the CE, in which $N_\pm$ are specified,
the normalization constant $Z_{L,N_+,N_-}^{\rm CE}$ 
is the sum of traces of matrix products over all
configurations which respect the given $N_+$ and $N_-$.  
This ensemble is in many ways the natural one; 
Monte Carlo simulations for example, are usually done in the CE. 
However, it seems difficult to compute
$Z_{L,N_+,N_-}^{\rm CE}$.  
It is easier to analyze the model
in a superposition of CE's, that is, in 
an appropriately defined GCE. 
There is a certain ambiguity in the definition of 
such an ensemble, but a standard
method has emerged \cite{DJLS,AHR98-4}. In the AHR model it would be natural
to introduce different fugacities $\xi_\pm$ for the two
types of particles, but since
we are interested in the case of equal densities, we set 
$\xi_+=\xi_-=\xi$.
Thus we take the probability of a configuration 
$\tau_1\,\tau_2\,\cdots\,\tau_L$,
where $\tau_j=+,-\,\, {\mbox {\rm or}}\,\, 0$ for $j=1,2,\ldots,L$, to be
\begin{equation}
  \label{mpa}
  P^{\rm GCE}(\tau_1,\tau_2,\ldots,\tau_L)
  =
  \frac{1}{Z_L(\xi)}
  \Tr
  \prod_{n=1}^{L}
  [ \delta_{\tau_n +}\xi D 
    +
   \delta_{\tau_n -} \xi E 
    +
   \delta_{\tau_n 0} A ] . 
\end{equation}
Now the normalization constant $Z_L(\xi)=Z_L^{\rm GCE}(\xi)$ is
a sum over traces of matrix products for 
{\it all} configurations of size $L$ which have
at least one hole, that is,
\begin{equation}
  \label{Z_L}
  Z_L(\xi)
  =
  \Tr'G^L
\end{equation}
with 
\begin{equation}
  G
  =
  A+\xi C
\end{equation}
where
\begin{equation}
  \label{G}
  C 
  = 
  D+E.
\end{equation}
Here the prime in $\Tr'$ indicates a sum over 
all terms with at least one $A$.
A more detailed discussion of the relationship between the
CE and the GCE will be given in section~5.

The current $J_+$ of $+$ particles in the GCE is defined
as the average flux of $+$ particles through a given bond.
It is the same as the current $J_-$ of $-$ particles,
defined similarly.
The current $J=J_+=J_-$ and the density 
$\rho=\rho_+=\rho_-$ are easily seen to be given by
\begin{align}
  \label{current_def}
  J_L(\xi)
  &=
  (1-q)\xi
  \frac{Z_{L-1}(\xi)}{Z_L(\xi)},
  \\
  \label{rho_def}
  \rho_L(\xi)
  &=
  \frac{\xi}{2L}
  \frac{\partial}{\partial \xi}
  \ln Z_L(\xi),
\end{align}
where the subscript $L$ indicates the system size.
As pointed out in the introduction, we are interested in 
the infinite volume limits 
$J(\xi)=\lim_{L\rightarrow\infty} J_L(\xi)$ and 
$\rho(\xi)=\lim_{L\rightarrow\infty} \rho_L(\xi)$ of 
these quantities because a phase transition between
the mixed and the disordered regime should manifest itself as
non-analyticity in $J$ as a function of $\rho$. $J(\xi)$ and 
$\rho(\xi)$ will be computed in the next section.

\Section{Exact Results in the Thermodynamic Limit}
Instead of considering $Z_L(\xi)$ directly,
we introduce the generating function
\begin{align}
  \label{Theta}
  \Theta(\lambda,\xi)
  =
  \sum_{L=1}^{\infty}
  \lambda^{L-1} Z_L(\xi).
\end{align}
This sum can be explicitly evaluated as follows.
 First we rewrite $Z_L(\xi)$ as a summation of the 
terms with the condition that the left-most hole is at
site $j+1$:
\begin{equation}
  \label{ZL}
  Z_L(\xi)
  =
  \sum_{j=0}^{L-1}
  \Tr (\xi C)^j A G^{L-j-1}
  =
  \sum_{j=0}^{L-1}
  \langle 0| G^{L-j-1} (\xi C)^j |0\rangle ,
\end{equation}
where we used the cyclicity of trace and 
the explicit representation of  $A=|0\rangle\langle 0|$.
Hence the generating function $\Theta(\lambda,\xi)$ is 
rewritten as
\begin{equation}
  \label{Theta_1}
  \Theta(\lambda,\xi)
  =
  \sum_{j,k=0}^{\infty}
  \langle 0| (\lambda G)^k (\lambda\xi C)^j |0\rangle .  
\end{equation}
Second we expand $G^k$ as
\begin{equation}
  \label{G_exp}
  G^k
  =
  \sum_{r\ge0}\hskip-30pt\sum_{
     \begin{array}{c}
      \scriptstyle j_0,j_r \geq 0 \\\noalign{\vskip-5pt}
      \scriptstyle j_1,m_1,\ldots,j_{r-1},m_r >0\\\noalign{\vskip-4pt}
      \scriptstyle j_0+m_1+j_1+\cdots + m_r+j_r = k
     \end{array}}
  (\xi C)^{j_0} A^{m_1} (\xi C)^{j_1} \cdots A^{m_r} (\xi C)^{j_r}.
\end{equation}
If we observe that $A^m=A$ and define
\begin{equation}
  \chi(x)
  =
  \sum_{n=0}^{\infty}
  \frac{1}{x^{n+1}}
  \langle 0|C^n|0\rangle
  = 
  \langle 0| \frac{1}{x-C} |0\rangle ,
\end{equation}
so that 
 \begin{equation}
  \sum_{n=0}^{\infty}(\xi\lambda)^n  \langle 0|C^n|0\rangle
    =\frac{1}{\xi\lambda}\chi\left(\frac{1}{\xi\lambda}\right), \qquad
  \sum_{n=0}^{\infty}(n+1)(\xi\lambda)^n   \langle 0|C^n|0\rangle
    =\frac{1}{\xi}\frac{d}{d\lambda}\chi\left(\frac{1}{\xi\lambda}\right),
 \end{equation}
then after some computation we obtain
\begin{equation}
  \label{Theta_exp}
  \Theta(\lambda,\xi)
  =
  \frac{  \displaystyle{\frac{d}{d \lambda}
          \chi \left( \frac{1}{\lambda\xi} \right) } }
       {  \xi -  \chi ( \frac{1}{\lambda\xi} )  }.
\end{equation}

 Finally it turns out that the function $\chi(x)$ has been 
known in mathematics literature \cite{AI84}. 
This is related to the fact that
the matrix $C$ is essentially the Jacobi matrix  
associated with certain $q$-orthogonal polynomials,
called Al-Salam-Chihara polynomials \cite{AC}
(and regarded as a special case of the Askey-Wilson polynomials
\cite{AW85}); see \cite{PASEP} for an explanation of this connection.
For example, from \cite{AI84}, equation (3.21), 
it follows  that 
\begin{equation}
  \label{chi}
  \chi(x)
  =
  f(y(x))
\end{equation}
with
\begin{align}
  \label{def_f}
  f(y) 
  &=
  y
  \frac{(q y^2;q)_{\infty}(q;q)_{\infty}}
       {(a y;q)_{\infty}^2}
  \sum_{n=0}^{\infty}
  \frac{(ay;q)_n^2 }
       {(qy^2;q)_n (q;q)_{n}} 
  q^n,
  \\
  \label{yx}
  y(x)
  &=
  \frac{x-2-\sqrt{x^2-4x}}{2},
\end{align}
where $(z;q)_n$ and $(z;q)_{\infty}$ are defined as
\begin{align}
  (z;q)_n
  &= \left\{
  \begin{array}{ll}1,& \text{if $n=0$},\\
  (1-z)(1-zq)(1-zq^2)\cdots(1-zq^{n-1}),& \text{if $n>0$},
   \end{array}\right.
  \\
  (z;q)_{\infty}
  &=
  \prod_{n=0}^{\infty} (1-zq^n).
\end{align}
If we introduce the basic hypergeometric function \cite{GR}
\begin{equation}
  \label{b_hyp}
  {_2\phi_1}
  \left[ \begin{array}{cc}  
               a_1,a_2  \\
               b       
         \end{array};
                        q ,z 
  \right]
  =
  \sum_{n=0}^{\infty}
  \frac{(a_1;q)_n (a_2;q)_n  }
       {(b;q)_n (q;q)_{n}} 
  z^n,  
\end{equation}
(\ref{def_f}) is rewritten as
\begin{equation}
  \label{f_byp}
  f(y) 
  =
  y
  \frac{(q y^2;q)_{\infty}(q;q)_{\infty}}
       {(a y;q)_{\infty}^2}
  {_2\phi_1}
  \left[ \begin{array}{cc}  
               ay,ay  \\
               qy^2       
         \end{array};
                        q ,q 
  \right].
\end{equation}
This is the explicit expression 
of the generating function $\Theta(\lambda,\xi)$.

Next we are interested in the asymptotic behavior of $Z_L(\xi)$ 
when $L\rightarrow\infty$. 
This is determined by the 
singularity closest to the origin of the generating function 
$\Theta(\lambda,\xi)$ as a function of $\lambda$.
We do not demonstrate this fact here; for a thorough discussion
see Appendix A of \cite{ERS}.
 From the expression (\ref{Theta_exp}), we see that 
there are two possible sources of singularities in 
$\Theta(\lambda,\xi)$: a singularity of the function 
$\chi\left(\frac{1}{\lambda\xi}\right)$ (the differentiation in the
numerator does not change the position
of any singularities) and a zero of the denominator.
As we will see, the singularity which is 
closest to the origin is always of the latter type,
for the region of the parameter space
of interest  here.

 First we consider the singularities of
$\chi\left(\frac{1}{\lambda\xi}\right)$. From (\ref{yx}) it follows that
$y(1/u)$ is analytic at $u=0$, that the singularity closest to the origin
is the square root singularity at $u=1/4$, 
and that $|y(1/u)|<1$ for $|u|<1/4$.
Moreover, from (\ref{def_f}), the singularity of $f(y)$ closest to the
origin is a simple pole at $y=1/a$.  Thus if $\lambda_0(\xi)$ denotes the
singularity of $\chi\left(\frac{1}{\lambda\xi}\right)$ closest to the
origin, then (i)~$\lambda_0(\xi) = 1/(4\xi)$ is a square root singularity
if $a<1$ and (ii)~$\lambda_0(\xi) = a/[(a+1)^2\xi]$ is a 
simple pole if $a>1$.
In what follows we restrict our detailed analysis to the case $a<1$. 
(See
remarks following (\ref{cn-aw}); the analysis of $a>1$ is similar.)

In order to discuss the zeros  of the denominator of (\ref{Theta_exp})
we make the following assumption:
\begin{equation}
\text{\bf  If $0\le q<1$ then the function $f$ 
    satisfies $f'(y)>0$ for $0\le y\le1$.}\label{assume}
\end{equation}
This assumption is key for our analysis; we have considerable 
evidence for its truth, which we will discuss
shortly, but at the moment no
proof.  From (\ref{chi}), (\ref{yx}), and (\ref{assume}) it follows that, 
when $\xi$ is positive, 
the function $\chi(\frac{1}{\lambda\xi})$ increases
monotonically from $0$ to $f(1)$ as $\lambda$ increases
from  $0$ to $\lambda_0(\xi)$.
Thus if $\xi$ satisfies 
$0 \leq \xi \leq \xi_{\text{max}}$, where $\xi_{\text{max}}=f(1)$, 
the equation
\begin{equation}
  \label{pole}
  \chi \left( \frac{1}{\lambda \xi} \right) 
  =
  \xi
\end{equation}
has a unique solution $\lambda(\xi)$ in the interval $[0,\lambda_0(\xi)]$,
and the denominator of (\ref{Theta_exp})
vanishes at this point. 
Since $\lambda(\xi)$ is smaller than $\lambda_0(\xi)$,
this is the singularity of the function 
$\Theta(\lambda,\xi)$ which is closest to the origin
when $0\leq \xi \leq \xi_{\text{max}}$.
This singularity is a simple pole, so that $Z_L(\xi)$ behaves
asymptotically ($L\rightarrow\infty$) as
\begin{equation}
  \label{ZL_largeL}
  Z_L(\xi)
  \simeq
  \text{const.}
  \left[ 
     \frac{1}{\lambda(\xi)}
  \right]^L
\end{equation}
Thus from (\ref{current_def}) and (\ref{rho_def}), 
\begin{align}
  J(\xi)
  &=\lim_{L\to\infty}J_L(\xi)
  =
  (1-q)\xi\lambda(\xi) ,
  \\
  \rho(\xi)
  &=\lim_{L\to\infty}\rho_L(\xi)
  =
  -\frac{\xi}{2} \frac{\partial}{\partial \xi} \ln \lambda(\xi).
\end{align}

These expressions can be simplified somewhat because
the function $y(x)$ is explicitly invertible.
If we denote the inverse function of $f(y)$ for $0\leq y \leq 1$ 
by $g(\xi)$,
\begin{equation}
  \label{fg}
  \xi = f(y)
  \Leftrightarrow
  y = f^{-1}(\xi) = g(\xi)
  \quad \text{for}  \quad 0\leq \xi \leq \xi_{\text{max}},\
   0\leq y \leq1,
\end{equation}
then the equation (\ref{pole}) can be  rewritten as
\begin{equation}
  \label{x_xiers}
  \lambda(\xi)
  =
  \frac{1}
       { \xi\chi^{-1}(\xi) }
  =
  \frac{g(\xi)}{\xi(1+g(\xi))^2}.
\end{equation}
Combining the above results, we find 
\begin{align}
  \label{J_g}
  J(\xi)
  &= 
  (1-q) 
  \frac{ g(\xi)}{(1+g(\xi))^2}, 
  \\
  \label{rho_g}
  \rho(\xi) 
  &= 
  \frac{1}{2}
  \left[
    1
    -
    \frac{\xi ( 1-g(\xi) ) g'(\xi)}{ (1+g(\xi))g(\xi)}
  \right] .
\end{align}
 From (\ref{assume})  and (\ref{fg}) it follows that 
$g(\xi)$ increases monotonically from 
$0$ to $1$ as $\xi$ increases from $0$ to $\xi_{\text{max}}$ and,
since $g'(\xi)=1/f'(g(\xi))$, that 
$0<g'(\xi)<\infty$ in this region.
Then (\ref{J_g})  and (\ref{rho_g}) imply that
$J$ and $\rho$ are analytic functions of $\xi$
throughout the range $0<\xi<\xi_{\text{max}}$,
and that as $\xi$ increases from $0$ to  $\xi_{\text{max}}$,
$J(\xi)$ increases monotonically from $0$ to $(1-q)/4$
and $\rho(\xi)$ increases from $0$ to $1/2$.
We expect also, and will assume in what follows, 
that $\rho(\xi)$ is a monotonic function throughout
the range $0\le\xi\le\xi_{\text{max}}$.
It follows that we may invert the function $\rho(\xi)$ as
$\xi(\rho)$ and thus 
obtain an analytic function $J(\rho)=J(\xi(\rho))$ defined  for 
$0\le\rho\le1/2$.  This is the indication that there is 
no mixed/disordered phase transition in
the GCE.

Note that for this argument it is important that $f'(y)$ 
be strictly positive not only for $0\leq y<1$ but also 
for $y=1$
(and hence $g'(\xi)$ finite not 
only for $0\leq \xi <\xi_{\text{max}}$ but 
also for $\xi=\xi_{\text{max}}$); 
this guarantees the convergence of $\rho(\xi)$ in (\ref{rho_g})
to $1/2$ as $\xi\rightarrow \xi_{\text{max}}$.

 To obtain $J(\rho)$ we must first 
invert the function 
$f(y)$ to obtain  $g(\xi)$, then
invert the function $\rho(\xi)$ 
to obtain the fugacity as a function of the density.
It is not possible to carry out these  inversions 
explicitly in general, and  we must  use numerical methods to 
obtain $J(\rho)$.
But there are special cases where one or 
both of the inversions can be done explicitly.
Let us study these cases first.

\vspace{5mm}
\noindent
{\it Case 1. $q=0$.} 
\nopagebreak
\noindent
 For this case, the formula for $f(y)$ in (\ref{def_f}) 
is greatly simplified since
$(z;0)_n=(z;0)_{\infty}=1-z$ and the only $n=0$ term
in the infinite series of (\ref{def_f}) remains.
We find 
\begin{align}
  \label{f_q0}
  f(y)  &= \frac{y}{(1-ay)^2},
  \\
  \label{fp_q0}
  f'(y) &= \frac{1+ay}{(1-ay)^3}.
\end{align}
Notice that the assumption (\ref{assume}) 
holds in this case, since $a>-1$.
The function $f(y)$ is invertible, and
\begin{equation}
  \label{q0g}
  g(\xi)
  =
  \frac{1+2a\xi-\sqrt{1+4a\xi}}
       {2 a^2 \xi}.
\end{equation}
The explicit formulae for $J(\xi)$ and $\rho(\xi)$ are then
given by
\begin{align}
  \label{q0J}
  J(\xi)
  &=
  \frac{2 a^2 \xi}
       {1+a^2+2a(1+a)^2\xi-(1-a^2)\sqrt{1+4a\xi}},
  \\
  \label{q0rho}
  \rho(\xi)
  &=
  \frac{a(1+a)\xi [(1+a)\sqrt{1+4a\xi}-(1-a)]}
       {\sqrt{1+4a \xi}[1+a^2+2a(1+a)^2\xi-(1-a^2)\sqrt{1+4a\xi}]}.
\end{align}

These formulae are further simplified when $a=0$ (as pointed out in the
introduction,  setting $a=q=0$ reduces  the AHR~model
to the two-species ASEP  studied in \cite{DJLS}).
Now  $f(y)=y$,  $g(\xi)=\xi$, and
\begin{align}
  \label{DJLS_J}
  J(\xi)
  &=
  \frac{\xi}
       {(1+\xi)^2},
  \\
  \label{DJLS_rho}
  \rho(\xi)
  &=
  \frac{\xi}
       {1+\xi}.
\end{align}
 From the easy inversion of  (\ref{DJLS_rho})  we obtain 
the known  current density relation
\begin{equation}
  \label{}
  J=\rho(1-\rho).
\end{equation}

\vspace{5mm}
\noindent
{\it Case 2. $a=-1$.} 
\nopagebreak
\noindent
This case corresponds to taking the limit $q\rightarrow 1$ with 
$\alpha$ fixed, and is thus strictly speaking 
outside our parameter region $0\leq q<1$, 
but it is nevertheless of interest.
Moreover, it   will play the  role of 
zeroth-order approximation in the analysis of the next section;
for this reason, the functions for this special case 
will be distinguished by the superscript $(0)$
in the following. 
Using the formula for ${_2\phi_1}$ \cite{GR},
\begin{equation}
  \label{phi_formula_0}
  {_2\phi_1}
  \left[ \begin{array}{cc}  
               a_1,a_2  \\
               b       
         \end{array};
                        q , \frac{b}{a_1 a_2}
  \right]
  =
  \frac{(b/a_1;q)_{\infty} (b/a_2,;q)_{\infty}}
       {(b    ;q)_{\infty} (\frac{b}{a_1 a_2};q)_{\infty}} ,
\end{equation}
we find 
\begin{align}
  \label{f_am1}
  f^{(0)}(y) = \frac{y}{(1+y)^2},
  \\
  \label{fp_am1}
  \frac{\partial}{\partial y}f^{(0)}(y) = \frac{1-y}{(1+y)^3}.
\end{align}
Assumption (\ref{assume}) is not strictly satisfied, since 
$\partial f^{(0)}(y)/\partial y =0$ for $y=1$, 
but $f(y)$ is strictly monotone for $0\le y\le1$ so that 
the derivation of the asymptotic form (\ref{ZL_largeL}) is still valid.
Note that (\ref{f_am1})  and (\ref{fp_am1}) are independent of $q$
and hence may be obtained by setting $a=-1$ 
in (\ref{f_q0}) and (\ref{fp_q0}).
Hence the explicit formulae for 
$g(\xi)$, $J(\xi)$ are obtained by simply taking $a=-1$ in 
(\ref{q0g}) and (\ref{q0J}):
\begin{align}
  \label{q1g}
  g^{(0)}(\xi)
  &=
  \frac{1-2\xi-\sqrt{1-4\xi}}
       {2 \xi},
  \\
  \label{q1J}
  J^{(0)}(\xi)
  &=
  \xi .
\end{align}
As for $\rho(\xi)$, 
we have $\rho^{(0)}(\xi)=0$ for $0\leq \xi < \xi_{\text{max}}=1/4$, while 
$\rho^{(0)}(1/4)$ is not well specified.
Thus  we cannot achieve nonzero density values by any specification of
the fugacity: the $q\rightarrow 1$ limit is a
very singular limit for the density-fugacity relation.

We now turn to the question of the validity of the key assumption
(\ref{assume}).  We believe, although
we have not proved,  that (\ref{assume}) holds 
for all $a,q$ satisfying  $-1<a<1$ and $0\leq q<1$.
This belief is based on several pieces of evidence.
We have evaluated the functions 
$f(y)$ and $f'(y)$ numerically for various values of $a$, $q$, and $y$
and have always found that $f'(y)>0$ for $0\le y\le1$  (such evaluations are
easily carried out to arbitrarily high accuracy from the formula
(\ref{def_f})).  We do observe, however, that 
for $q$ and $y$ close to 1
the value of $f'(y)$ is very small;
for instance, $f'(1)$ is of order $10^{-36}$ 
when $q=0.9$ and $\alpha=1$.  This phenomenon will
play a key role in the analysis of section~4.
Note also that it follows from the considerations of the 
special cases above that (\ref{assume}) holds when $q=0$, and that 
when $a=-1$ (recall that this corresponds to a limit $q\to1$),
$f'(y)>0$ when $0\leq y <1$; this suggests that $f'(y)>0$ 
except at the corner $q=y=1$ of the region of interest.
Moreover, we will show in the next section 
that $ f'(1) > 0 $ when $\alpha=1$ (i.e., $q=-a$).
This is a weaker statement than (\ref{assume})
but, since $f(y)$ is expected to be a monotonically decreasing
function, is a strong analytical indication that
(\ref{assume}) holds when $q=-a$.

\Section{Two Distinct Regimes without Non-analyticity}

In the previous section we argued that, even in infinite volume,
the current is an analytic function of the 
density throughout the mixed/disordered phase. 
On the other hand, it is clear from inspection of 
the $J$-$\rho$ diagram in
the thermodynamic limit that, for large values of $q$,
the derivative of
$J(\rho)$ {\it does} change rapidly---in fact, 
extraordinarily rapidly---as
$\rho$ passes through some critical value $\rho_c$.  
For example, when $q=-a=0.9$, there is a change of 
order one in $J'(\rho)$ arising from a
change of order $10^{-24}$ in $\rho$ (see Fig.~2).  
In this section, we
explain how this behavior arises: it is 
associated with the very small
values of $f'(1)$ occurring when $q$ is near $1$, 
which lead to rather
peculiar behavior in the density-fugacity relation (\ref{rho_g}).  

In Fig.~3, we graph the functions $J(\xi)$ 
and $\rho(\xi)$ of (\ref{J_g}),
(\ref{rho_g}) for $\alpha=1$ and several values of $q$.  
Note that, although the function $J(\xi)$ 
remains quite smooth as $q$ increases, and indeed even for $q=1$, 
the function $\rho(\xi)$ develops an apparently
sharp corner or kink, 
even for values of $q$ about $0.75$; this kink corresponds to the 
kink in the $J$-$\rho$ curve at $\rho_c$.  Here we
give an approximate analysis of this behavior under the assumption 
that $q$ is reasonably near to 1 and that $f'(1)$ is very small. 

We first analyze the region $\rho<\rho_c$.
Using the formula (see \cite{GR})
\begin{equation}
  \label{phi_formula}
  {_2\phi_1}
  \left[ \begin{array}{cc}  
               a_1,a_2  \\
               b       
         \end{array};
                        q ,z 
  \right]
  =
  \frac{(a_2;q)_{\infty}(a_1z;q)_{\infty}}
       {(b;q)_{\infty}(z;q)_{\infty}}
  {_2\phi_1}
  \left[ \begin{array}{cc}  
               b/a_2,z  \\
               a_1 z       
         \end{array};
                        q, a_2 
  \right],  
\end{equation}
we obtain an alternate expression for $f(y)$:
\begin{equation}
 \label{f_simple}
  f(y)
  =
  \frac{y}{1-ay}
  \sum_{n=0}^{\infty}
  \frac{(a^{-1}qy;q)_{n}}
       {(aqy;q)_{n}} 
  (a y)^n . 
\end{equation}
Since we are interested in values of $q$ close to 1 
we set $q=1-\epsilon$, where  
$\epsilon(>0)$ is assumed to be small, and expand in $\epsilon$,
denoting quantities accurate to order $k$ by a corresponding
superscript (e.g. $f^{(k)}$).
The $0^{\rm th}$ order was analyzed in the previous section; here
we will carry out the expansion to first order. 
By a straight-forward computation we obtain
\begin{equation}
  \label{app_f}
  f^{(1)}(y)
  =
  \frac{y}{(1+y)^2}
  \left[
  1
  +
  \frac{2y}{\alpha(1+y)^2} 
  \epsilon
  \right]
  = f^{(0)}(y)\left[1+\frac{2f^{(0}(y)}{\alpha}\epsilon\right].
\end{equation}
As $y$ increases from zero to one,
$f^{(1)}(y)$ increases monotonically from zero
to $\xi_{\text{max}}^{(1)}=1/4 + \epsilon/(8 \alpha)$.
The function $f^{(1)}(y)$ is invertible, with 
inverse
\begin{equation}
  \label{g_q}
  g^{(1)}(\xi)
  =
  g^{(0)}(X(\xi))
  =
  \frac{1-2X(\xi)-\sqrt{1-4X(\xi)}}{2X(\xi)},
\end{equation}
where
\begin{equation}
  X(\xi)
  =
  \frac{\alpha}{4 \epsilon}
  \left( \sqrt{1+\frac{8\epsilon\xi}{\alpha}} -1 \right),
\end{equation}
for $0\leq \xi \leq \xi_{\text{max}}^{(1)}$. Then
the density-fugacity relation (\ref{rho_g}) becomes
\begin{equation}
  \label{den_fu}
  \rho(\xi)
  =
  \frac{1}{2}
  \left[  1-\frac{\xi \frac{d}{d\xi}X(\xi) }{X(\xi)} \right] + O(\epsilon^2),
\end{equation}
so that to first order in $\epsilon$ the density is
\begin{equation}
  \rho^{(1)}(\xi)
  = \frac{\epsilon \xi}{\alpha}.
\end{equation}
Comparing these results with the curve
for $q=0.75$ in Fig.~3,
we realize that (\ref{den_fu}) gives the behavior 
of the density-fugacity relation up to the kink;
since this region includes 
$\xi=0$ (or $\rho=0$), it corresponds to 
the disordered regime. 
As the fugacity $\xi$ increases from zero to $\xi_{\text{max}}^{(1)}$, 
the density $\rho^{(1)}$ increases linearly from zero to
$\epsilon/(4\alpha)$, so that this value may be identified,
up to first order, with the density $\rho_c$:
\begin{equation} 
\label{rhoc}
\rho^{(1)}_c
=
\frac{\epsilon}{4\alpha}.
\end{equation}

Note that under this analysis 
there is no value of $\xi$ which gives rise to a density 
$\rho$ in the interval  $(\rho_c,\frac12]$.
This is because our approximation of $f(y)$ by 
$f^{(1)}(y)$  lacks an important property of 
the original $f(y)$: 
the derivative of $f^{(1)}$ becomes zero at $y=1$,
whereas our basic assumption (\ref{assume}) is that $f'(1)>0$. 
In fact, it appears that $f'(1)$ vanishes to all orders in the 
perturbation expansion in $\epsilon$; this is shown below
(see (\ref{fp1_asymp})) in the case $q=-a$.

We now analyze the region 
$\rho_c<\rho\le\frac12$ under the assumption 
that $f'(1)$ is positive but very small.
We will also assume (as is supported
by numerical evaluations) that $f''(1)$ is negative and of order unity.
Under these assumptions,
and recalling that $f(1)=\xi_{\text{max}}$,
we may approximate $f(y)$ near $y=1$ as
\begin{equation}
  \label{local_f}
  f(y)
  \simeq f^*(y)
  = 
  \xi_{\text{max}}
  -
  B ((1+\delta-y)^2-\delta^2),
\end{equation}
where $B=-f''(1)$ and $\delta=-f'(1)/f''(1)$; that is, 
$f(y)$ will have 
a quadratic maximum at approximately $1+\delta$, 
where $\delta$ is very small.
The function $f^*(y)$ has inverse
\begin{equation}
  \label{}
  g^{*}(\xi)
  =
  1+\delta-\sqrt{\frac{\xi_{\text{max}}-\xi}
                      {B}                    +\delta^2}
 =
  1+\delta-\delta\sqrt{\eta+1},
\end{equation}
where for $\xi\simeq\xi_{\text{max}}$,
it is convenient to introduce the scaled variable $\eta$ defined by
 \begin{equation}
 \label{eta}
 \xi_{\text{max}}-\xi=\eta \delta^2.
 \end{equation}
The density-fugacity relation (\ref{rho_g}) then becomes 
$\rho(\xi)\simeq\rho^*(\xi)$, where 
\begin{equation}
  \label{den_eta}
  \rho^*(\xi)
  =
  \frac12
  \left[ 
     1-\frac{\xi_{\text{max}}}{4B}
       \frac{\sqrt{\eta+1}-1}{\sqrt{\eta+1}} 
  \right],
\end{equation}
so that as $\eta$ goes infinity, $\rho^*(\xi)$ approaches
\begin{equation}
  \label{rhocstar}
  \rho_c^{*}
  =
  \frac12
  \left[
     1-\frac{\xi_{\text{max}}}{4 B}
  \right]. 
\end{equation}
Thus here the critical density $\rho_c$ arises as the minimum value
of the density which can be described by  (\ref{den_eta}). 
Clearly (\ref{den_eta}) describes the density-fugacity relation
in the mixed regime, that is, for $\rho>\rho_c$.
When $q=0.75$, for example, this corresponds 
to the nearly vertical line shown in Fig.~3.
The value $\rho_c$ of (\ref{rhocstar}) is an extremely accurate 
value for the  ``critical density'' between the mixed and
disordered regimes when $q$ is close to one; this is the value used in
plotting Fig.~2 and thus in the case $q=-a=0.9$ is accurate to about 
$10^{-24}$.

Note that (\ref{app_f}) implies that to
first order in $\epsilon$, $B=1/16+\epsilon/(16 \alpha)$; recalling that 
$\xi_{\text{max}}^{(1)}=1/4 + \epsilon/(8 \alpha)$, we see 
that (\ref{rhocstar}) reduces to (\ref{rhoc}) in first order.  In 
\cite{AHR98-4} the critical value of $\rho$ was estimated, on the basis of 
extrapolation of finite size data, to be (in our notation)
$(1-q)/[4\alpha-2(1-q)]$; this value agrees with (\ref{rhocstar}) to first
order in $\epsilon$.

The key point for the above analysis, as we have 
emphasized, is that $f'(1)$ is ``very small''
over a range of $q$ values close, but not necessarily very close, to 1. 
We can throw some additional light on this phenomenon because,
in the  special case $\alpha=1$,
it is possible to compute $f'(1)$ exactly.
When $\alpha=1$, i.e., when $a=-q$,  
the expression of $f(y)$ in (\ref{f_simple}) can 
be further simplified as
\begin{equation}
  \label{simple_f}
  f(y)
  =
  \frac{1}{1-q}
  \left[
     (1+y)^2 
     \sum_{n=0}^{\infty} 
     \frac{(-qy)^n}{1+yq^n}
     -1
  \right],
\end{equation}
 so that
\begin{equation}
  \label{fp1}
  f'(1)
  =
  \frac{4}{1-q}
  \left[
     \sum_{n=0}^{\infty}
     \frac{(-q)^n}{(1+q^n)^2}
     +
     \sum_{n=1}^{\infty}
     \frac{n (-q)^n}{1+q^n}
  \right]
\end{equation}
The sums in (\ref{fp1}) can be evaluated explicitly,
using the formulae \cite{GR}
\begin{equation}
  \label{formula}
  \sum_{n=1}^{\infty}
  \frac{(-q)^n}{(1+q^n)^2}  
  =
  \sum_{n=1}^{\infty}
  \frac{n (-q)^n}{1+q^n}
  =
  \frac{1}{8}
  \left(
     \left[
        \frac{(q;q)_{\infty}}{(-q;q)_{\infty}}
     \right]^4
     -1
  \right).
\end{equation}
This leads to
\begin{equation}
  \label{fp1_ans}
  f'(1)
  =
  \frac{1}{1-q}
  \left[
     \frac{(q;q)_{\infty}}{(-q;q)_{\infty}}
  \right]^4 .  
\end{equation}
This expression is valid for any value of $q$ 
satisfying $0 \leq q <1$, and is clearly positive.
The behavior of $f'(1)$ as a function of $q$
is shown in Fig.~4.

Now we consider the asymptotic
behavior of $f'(1)$ as $\epsilon = 1-q$ becomes small.
Setting $q=e^{-\pi^2 t}$, we find
\begin{equation}
  \label{q_th}
  \frac{(q;q)_{\infty}}
       {(-q;q)_{\infty}}
  =
  \prod_{n=1}^{\infty}
  \frac{1-q^n}{1+q^n}
  =
  \sum_{n=-\infty}^{\infty}
  (-)^n e^{- \pi^2 n^2 t} 
  =
  \vartheta_0(0,i \pi t).
\end{equation}
Here $\vartheta_0(x,\tau)$ is an elliptic theta function 
defined by
\begin{equation}
  \label{theta0}
  \vartheta_0(x,\tau)
  =
  \sum_{n=-\infty}^{\infty}
  (-)^n e^{i \pi \tau  n^2 + 2\pi i x n}.
\end{equation}
Applying Jacobi's imaginary transformation,
\begin{equation}
  \label{jacobi}
  \vartheta_0(x,\tau)
  =
  \frac{1}{\sqrt{-i \tau}}
  e^{-i\pi x^2/\tau}
  \vartheta_2 \left( \frac{x}{\tau},-\frac{1}{\tau} \right),
\end{equation}
with 
\begin{equation}
  \label{theta2}
  \vartheta_2(x,\tau)
  =
  \sum_{n=-\infty}^{\infty}
  e^{i \pi \tau  (n-1/2)^2 + i\pi x (2n-1)},
\end{equation}
we obtain
\begin{equation}
  \label{final}
  \frac{(q;q)_{\infty}}
       {(-q;q)_{\infty}}
  =
  \frac{1}{\sqrt{\pi t}}
  \sum_{n=-\infty}^{\infty}
  e^{- (n-1/2)^2/t }.
\end{equation}
(These formula involving theta  functions are easily obtained from, for
example, Chapter 21 of \cite{WW}.) 
Now it is easy to read off the asymptotic behavior as 
$q$ approaches $1$. The leading contributions
come from the $n=0$ and $n=1$ terms in (\ref{final}). 
Since $t \simeq \epsilon/\pi^2+\epsilon^2/(2\pi^2)$, we find
\begin{equation}
  \frac{(q;q)_{\infty}}
       {(-q;q)_{\infty}}
  \simeq
  \frac{2}{\sqrt{\pi t}}
  e^{-1/4t}
  \simeq
  2\sqrt{\frac{\pi}{\epsilon}}
  \,e^{\pi^2\over8}\,
  e^{-\pi^2/4\epsilon}.
\end{equation}
 Finally we obtain the asymptotic expression for $f'(1)$:
\begin{equation}
  \label{fp1_asymp}
  f'(1)
  \simeq
  \frac{2^4\pi^2}{\epsilon^3} 
  \,e^{\pi^2\over2}\,
  e^{-\pi^2/\epsilon}.
\end{equation}
Using (\ref{rho_g}), it is also possible to obtain
the asymptotic expression for $\delta$:
\begin{equation}
 \delta = \frac{f'(1)}{2B}
 \simeq\frac{2^7\pi^2}{\epsilon^3}
 \,e^{\pi^2\over2}\,
 e^{-\pi^2/\epsilon}.
\end{equation}

\Section{Density Fluctuations and the Canonical Ensemble}

In the previous two sections,  we have studied 
the current in the thermodynamic limit
in the framework of a GCE. 
Our analysis strongly suggests that there is no
phase transition between the mixed and
disordered regimes in the GCE.
In this section we argue that in the thermodynamic
limit, the current in the 
GCE and the CE are the same, that is, that the current $J(\rho)$
discussed in Section 3 and 4 is the same as the infinite volume
limit of the current in the CE ensemble with density $\rho$.
This implies that the lack of a phase transition in the GCE 
discussed above also holds for the CE.  
The argument also sheds some light on
the structure of the infinite volume state of the model in the
mixed/disordered regime. 

In finite volume, the GCE is a superposition of canonical ensembles:  
contributions come from all 
values of the densities, with weights 
which are determined by the fugacity $\xi$.
The total mean density of particles (both $+$ and $-$) 
is $2\rho_L(\xi)$, where
$\rho_L(\xi)$ is given by formula (\ref{rho_def}).
The fluctuations in the total density can be expressed  
similarly (ideally we should discuss fluctuations in the density of each
species, but that seems more difficult).
If we write $N=N_++N_-$ and introduce the quantity
\begin{equation}
 \label{F_L}
  F_L(\xi)
  =
  {1\over L}\left(\E{N^2}_L-\E{N}_L^2\right)
  =
  {1\over L}\left(\xi{\partial\over\partial\xi}\right)^2
  \ln Z_L(\xi),
\end{equation}
then the fluctuations in the total density are simply given by $F_L(\xi)/L$.
Now if we take the thermodynamic limit in the GCE 
for some fixed $\xi$, then 
the mean total density $2\rho_L(\xi)$ is expected to have a well defined 
limiting value, 
$2\rho(\xi)=\lim_{L\rightarrow\infty}2\rho_L(\xi)$.  
If we can show that the fluctuations in the total density go to zero 
in the thermodynamic limit, then we expect that in this limit 
only a single total density $2\rho(\xi)$ survives, and  if we  ignore 
the possibility
of canceling fluctuations in the $+$ and $-$ densities then these densities
will also be unique in the limit. Thus,
in the thermodynamic limit, 
the GCE with fugacity $\xi$ is the same as 
a CE with ($+$ or $-$) density $\rho(\xi)$. 
This corresponds to the usual GCE/CE equivalence for equilibrium
systems away from a phase transition.
For our model, of course, condensation of particles in the mixed regime
suggests a possible phase transition, and we need an independent 
argument.
It turns out that we can exactly calculate  
$F(\xi)=\lim_{L\to\infty}F_L(\xi)$ from (\ref{rho_g}):
\begin{align} 
  F(\xi)
  &=2\xi\frac{\partial}{\partial\xi}\rho(\xi)
 ={\xi\over g(\xi)^2[1+g(\xi)]^2}\Bigl(\xi(1+2g(\xi)-g(\xi)^2)g'(\xi)^2
 \notag\\
  &\hskip150pt-\xi(1-g(\xi)^2)g(\xi)g''(\xi)-(1-g(\xi)^2)g(\xi)g'(\xi)\Bigr).
\end{align} 
Numerical values of this quantity can be very large.
For instance, for $q=-a=0.9$, we find that $F(\xi)$ is of order $10^{70}$
in the region 
$\rho>\rho_c$. However, the important point is that 
$F_L(\xi)$ has a finite value 
in the thermodynamic limit, so that the 
density fluctuations $F_L/L$ in the finite system vanish
in the infinite system. As argued above, this suggests 
that our analysis in the previous two sections,
of the behavior of $J(\rho)$ in 
the GCE, is valid for the CE as well.  

It is also possible to estimate the length scale $R$ up 
to which particle condensation will be observed
in Monte Carlo simulations.  Let us write 
$\E{N^2}=\sum_{1\le i,j\le L}\E{\sigma_i\sigma_j}$ in 
(\ref{F_L}), where $\sigma_i=1$ if there is a
particle at site $i$ and $\sigma_i=0$ otherwise.
If we assume that the particle-particle correlation function decays
exponentially with some characteristic length $R$,
$\E{\sigma_i\sigma_j}\sim e^{-|i-j|/R}$, then we see that $F_L\sim R$.
Thus the correlation length can become huge;
for instance, for the above case of $q=-a=0.9$, $R$ would be of 
the order $10^{70}$. Therefore it is certainly not possible
to observe the breakdown of the particle condensation with 
Monte Carlo simulations.

The discussion of this section also throws light on our assumption of
section~3, that $\rho(\xi)$ is a strictly increasing function.  Since
$2\xi\partial\rho_L/\partial\xi=F_L$ is a fluctuation,
$\partial\rho_L/\partial\xi$ must be positive, so that
$\partial\rho/\partial\xi\ge0$. Vanishing of $\partial\rho/\partial\xi$
could occur only with subnormal fluctuations in $N$
($\lim_{L\to\infty}F_L=0$), a scenario which is certainly unlikely and has
been ruled out in certain equilibrium situations
\cite{Ginebre}.

\Section{Concluding Remarks}

In this article we have studied an exclusion process on a ring,
originally introduced and investigated by Arndt {\it et al} 
\cite{AHR98-3,AHR98-4}. Using the theory of $q$-orthogonal polynomials,
we have obtained the exact expression of the current 
in the thermodynamic limit in the framework of a grand 
canonical ensemble (GCE). Contrary to what is suggested by Monte Carlo 
simulations and finite volume
calculations,\cite{AHR98-3,AHR98-4}, we find 
no phase transition between the mixed and the disordered regime;
specifically, the infinite volume current as a function of the density 
is analytic through the mixed/disordered phase.
However, the derivative of the current
can change rapidly over a {\it very} small (but finite) interval of $\rho$,
especially if $q$ is close to $1$. 

Our analysis of the infinite volume current depends on
a key assumption, inequality (\ref{assume}).  Although 
this is unproved, we have strong
evidence of its truth: it holds in the special case $q=0$ (and essentially
also the special case $a = -1$); it is supported by the analysis
of section~4 for the case $q=-a$; 
and we have checked its validity for various parameter 
values by highly accurate numerical computations.  However,
the rigorous proof of (\ref{assume}) remains an outstanding
problem.  A second assumption, the positivity of $\rho'(\xi)$, is physically
natural and is also supported by numerical calculation. 

Most of our analysis was carried out in a GCE.  The relation between the
canonical and grand canonical ensembles in equilibrium statistical
mechanics has been well studied, but we know of no general results for
non-equilibrium situations.  Here we have simply asked how far the results
for our GCE (which itself represents a particular and perhaps non-canonical
choice) are also valid for the CE. 
We have given a heuristic argument: exact calculations show that 
the density fluctuations in the GCE
vanish in the thermodynamic limit, which 
suggests that, in this limit, the GCE for a given fugacity agrees 
with the CE for a certain corresponding density.

It is fascinating that in the relatively simple AHR model one 
already has a quite subtle transition between the disordered and
the mixed regime. As discussed in section~5, computer simulations would
need to be carried out up to lattice sizes of the order $10^{70}$
(when $q=-a=0.9$)
to see the breakdown of the particle condensation, so that 
this is certainly a phenomenon which one 
would not expect from the accessible finite size Monte Carlo data.

\section*{Acknowledgment}
The authors would like to thank J. L. Lebowitz 
for fruitful discussions and comments. 
NR gratefully acknowledges a postdoctoral fellowship from the Deutsche
Forschungsgemeinschaft and thanks Joel Lebowitz for hospitality at the
Mathematics Department of Rutgers University and for support under NSF
grant DMR~95--23266, and thanks DIMACS and its
supporting agencies, the NSF under contract STC--91--19999 and the NJ
Commission on Science and Technology, for support. 
TS thanks the continuous
encouragement of M. Wadati. 
TS is a Research Fellow of the Japan Society
for the Promotion of Science.


\newpage
\begin{large}
\noindent
Figure Captions
\end{large}

\vspace{10mm}
\noindent
Fig. 1: 
Space time diagrams of the AHR~model from Monte Carlo
simulations. The horizontal axis represents the site number $j$ 
whereas the vertical axis represents time.
The existence of a positive (resp. negative) 
particle is represented as a black (resp. gray) point.
 For all three figures, we set 
$L=100,\, N_+=N_-=30, \, \alpha=0.5$.
The left-most figure corresponds to the pure phase
($p_R=0.9,p_L=1$).
The figure in the middle corresponds to the 
mixed regime ($p_R=1,p_L=0.9$).
The right-most figure corresponds to the disordered regime
($p_P=1,p_L=0$).

\vspace{10mm}
\noindent
Fig. 2 : 
The $J$-$\rho$ diagram for the case $\alpha=1,q=0.9$.
The horizontal and vertical axes of the inset are
$(\rho-\rho_c)\times 10^{24}$ and 
$(J/(1-q)-0.25)\times 10^{23}$ respectively.  
The function $J(\rho)$ is analytic,
although its derivative changes very rapidly as $\rho$ passes $\rho_c$.

\vspace{10mm}
\noindent
Fig. 3 : 
The functions $J(\xi)$ and $\rho(\xi)$ when $\alpha=1$.
For $q=0$, we have $J(\xi)=\xi/(1+\xi)^2$ (\ref{DJLS_J})
and $\rho(\xi)=\xi/(1+\xi)$ (\ref{DJLS_rho}). 
For $q=1$, we have $J(\xi)=\xi$ (\ref{q1J})
and $\rho(\xi)=0 \,(0\leq \xi < 1/4)$.
The function $J(\xi)$ for $0<q<1$ interpolates
these two limiting cases smoothly. 
On the other hand, 
as $q$ increases from zero, there appear two distinct 
regions for $\rho(\xi)$. 
In particular, the $q\rightarrow 1$ limit appears to be singular.

\vspace{10mm}
\noindent
Fig. 4 : 
The behavior of $f'(1)$ as a function of $q$.
We see that $f'(1)$ becomes extremely small
as $q$ approaches one.


\newpage
\renewcommand{\thepage}{Figure 1}
\begin{picture}(500,600)
\put(0,200){\scalebox{0.6}
{\includegraphics{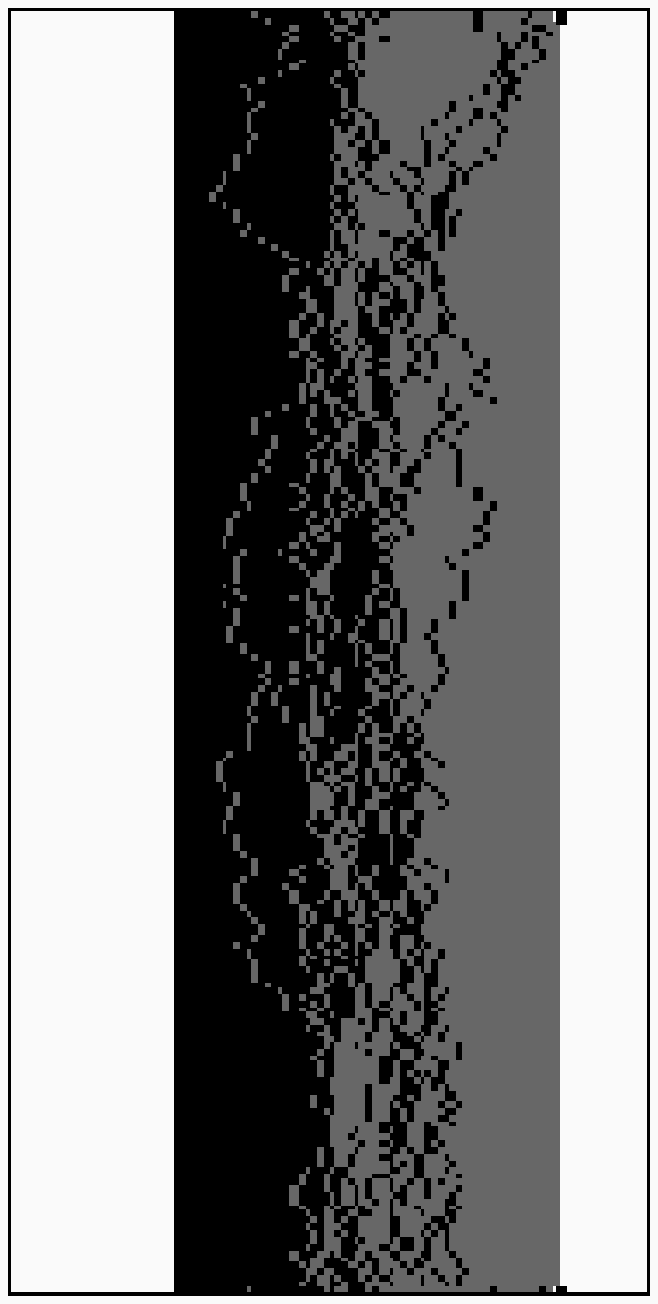}}}
\put(36,195){\vector(1,0){50}}
\put(60,177){$j$}
\put(-10,300){\vector(0,1){70}}
\put(-27,330){\rotatebox{90}{$t$}}

\put(150,200){\scalebox{0.6}
{\includegraphics{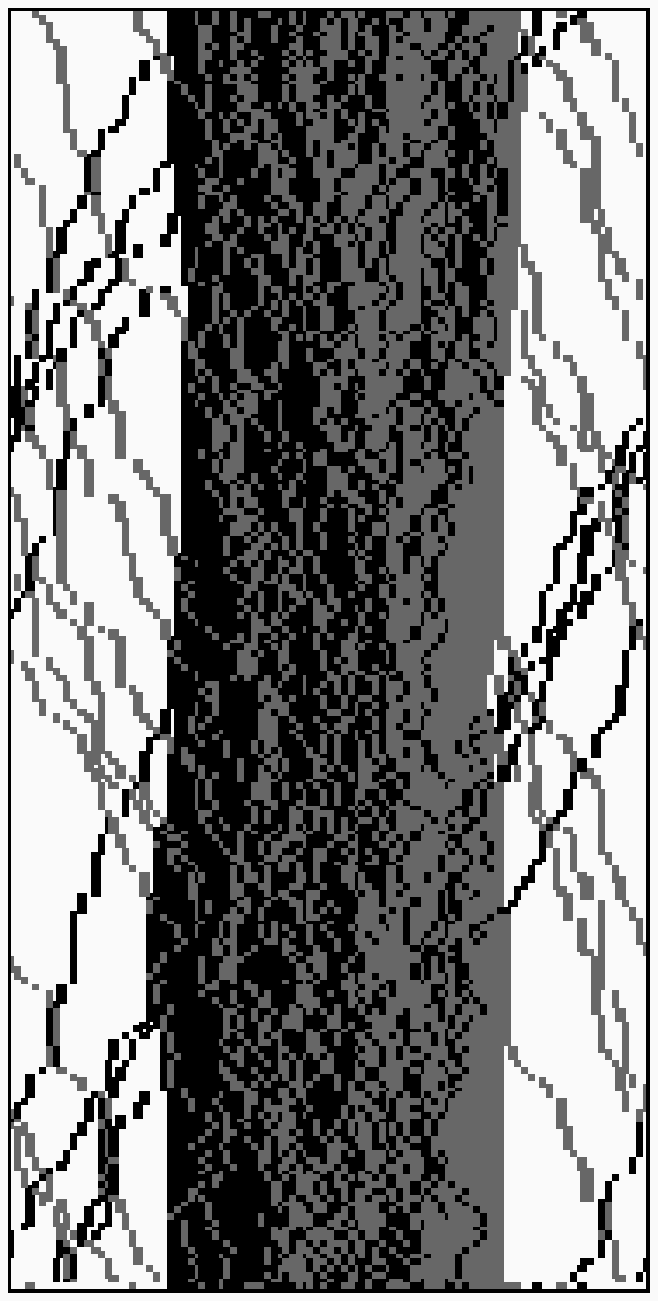}}}
\put(186,195){\vector(1,0){50}}
\put(210,177){$j$}

\put(300,200){\scalebox{0.6}
{\includegraphics{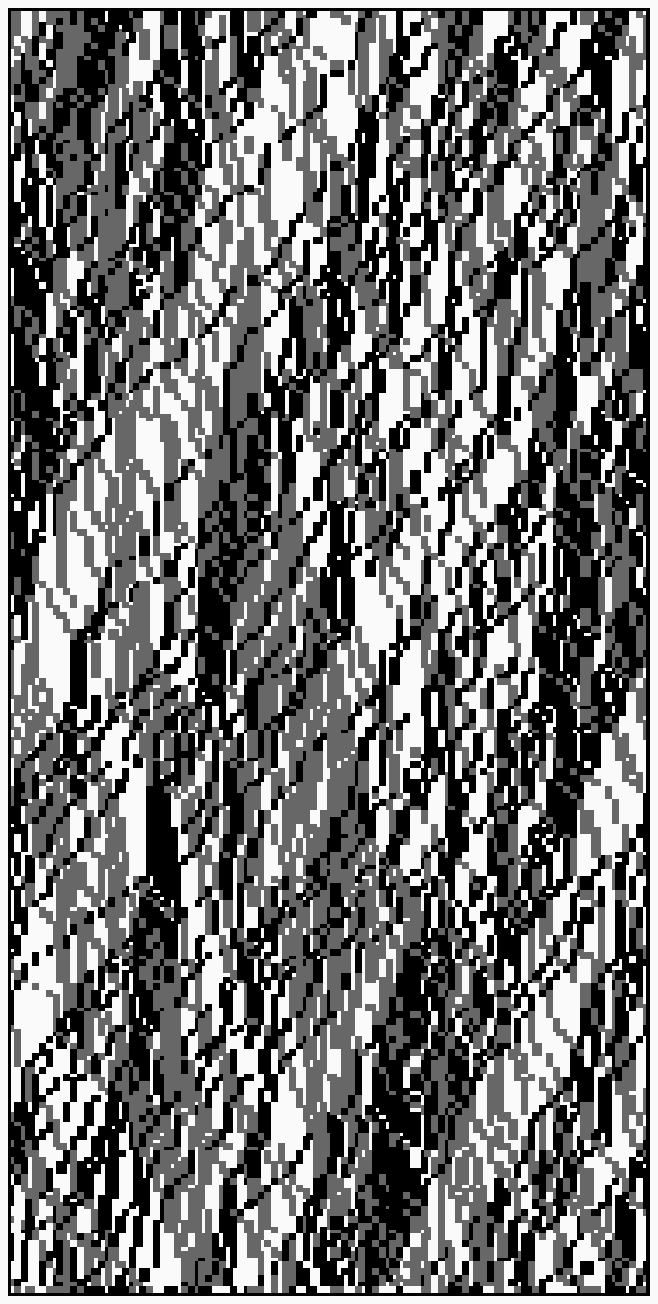}}}
\put(336,195){\vector(1,0){50}}
\put(360,177){$j$}

\end{picture}

\newpage 
\renewcommand{\thepage}{Figure 2}
\begin{picture}(500,500)
\put(20,100){\scalebox{1.5}
{\includegraphics{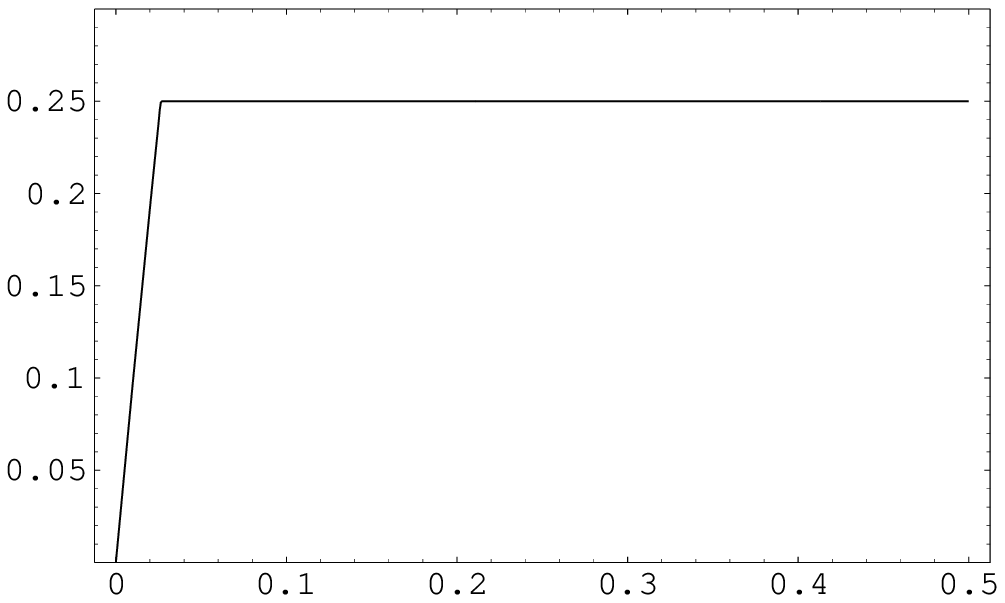}}
}
\put(265,70){\scalebox{1.5}{$\rho$}}
\put(-40,260){\scalebox{1.5}{$\displaystyle\frac{J}{1-q}$}}
\put(137,130){\scalebox{0.9}
{\includegraphics{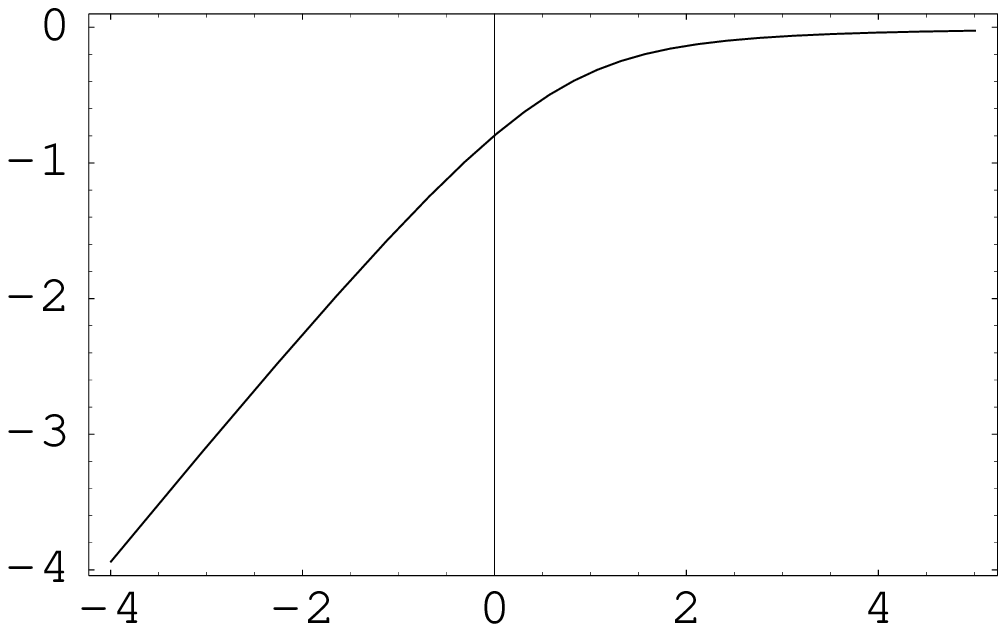}}
}
\put(130,290){\vector(-3,2){30}}
\put(375,130){\scalebox{1.}{$(\rho-\rho_c)\!\times\! 10^{24}$}}
\put(90,260){\scalebox{1.0}{$(\frac{J}{1-q}-\frac{1}{4})$}}
\put(95,245){\scalebox{1.}{$\times\! 10^{23}$}}
\end{picture}

\newpage
\renewcommand{\thepage}{Figure 3}
\begin{picture}(400,300)
\put(0,100){\includegraphics{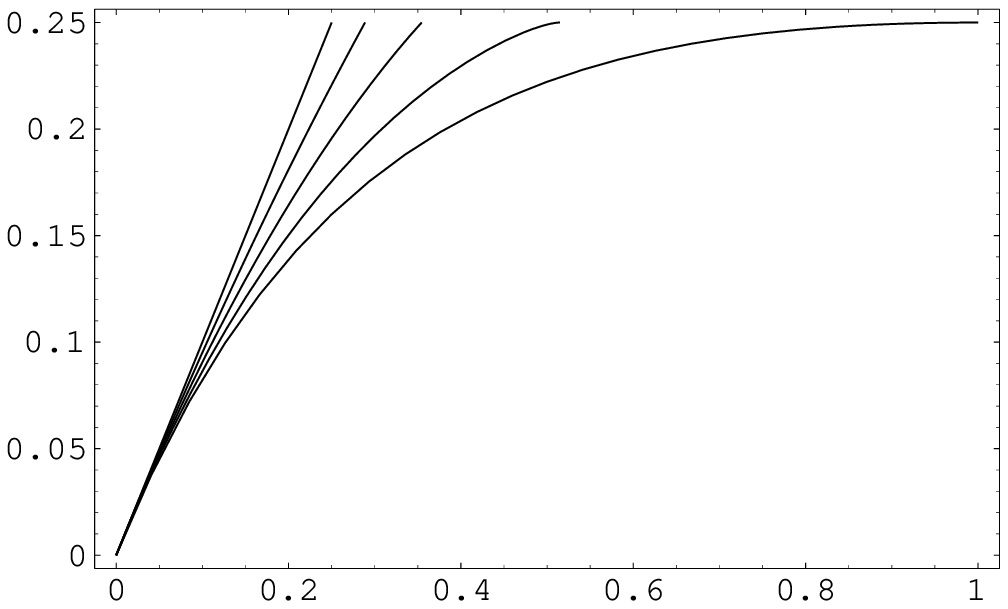}}
\put(160,80){$\xi$}
\put(-30,210){$J(\xi)$}
\put(238,300){$q=0$}
\put(250,290){\vector(0,-1){16}}
\put(170,300){$q=0.25$}
\put(180,290){\vector(-1,-1){16}}
\put(120,300){$q=0.5$}
\put(130,290){\vector(-1,-2){8}}
\put(70,300){$q=0.75$}
\put(97,290){\vector(1,-2){8}}
\put(37,258){$q=1$}
\put(70,260){\vector(1,0){15}}
\end{picture}


\begin{picture}(400,300)
\put(0,100){\includegraphics{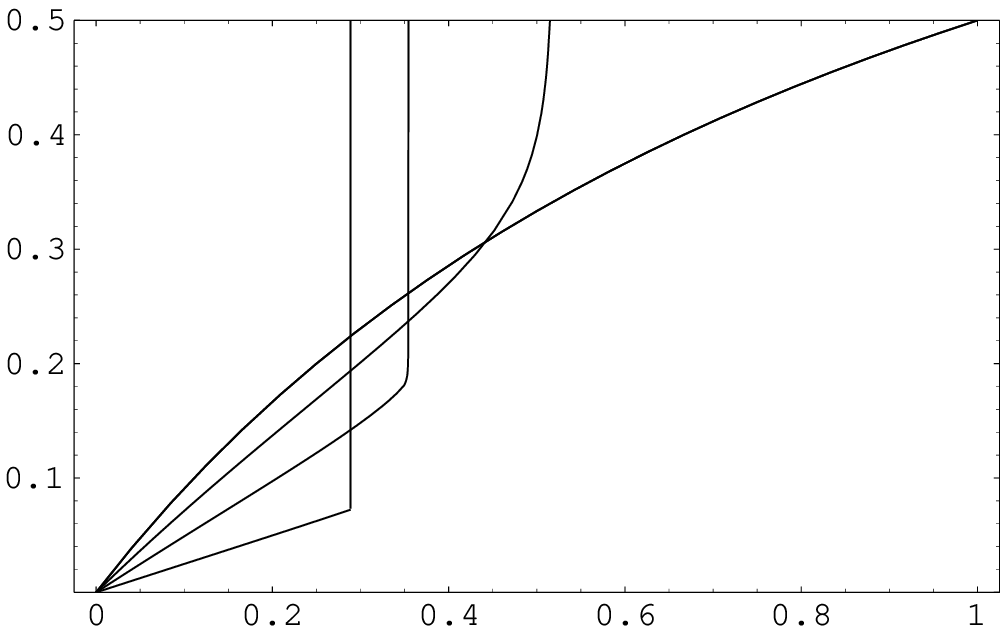}}
\put(160,80){$\xi$}
\put(-30,210){$\rho(\xi)$}
\put(238,310){$q=0$}
\put(250,300){\vector(0,-1){28}}
\put(174,310){$q=0.25$}
\put(177,300){\vector(-1,-1){16}}
\put(120,310){$q=0.5$}
\put(127,300){\vector(-1,-2){8}}
\put(60,310){$q=0.75$}
\put(80,300){\vector(1,-1){16}}
\end{picture}

\newpage
\renewcommand{\thepage}{Figure 4}
\begin{picture}(400,300)
\put(50,100){\includegraphics{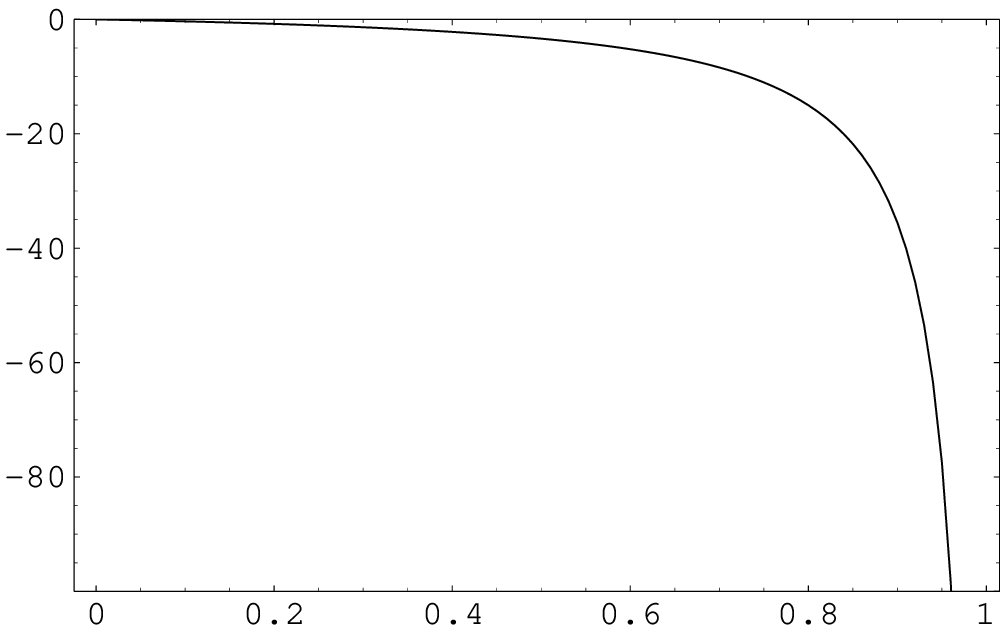}}
\put(210,80){$q$}
\put(-5,210){log$_{10} f'(1)$}
\end{picture}

\end{document}